%% file: apssamp.tex
\documentclass[%
 preprint, 
 amsmath,amssymb,
 aps, physrev,
]{revtex4-2}

\usepackage[table]{xcolor}
\usepackage{graphicx}
\usepackage{dcolumn}
\usepackage{dsfont}
\usepackage{enumitem}

\usepackage{hyperref}
\usepackage{cleveref}
\usepackage{tikz}
\usepackage{bbm}
\usetikzlibrary{tikzmark}
\usepackage{pgfplots}
\DeclareUnicodeCharacter{2212}{−}
\usepgfplotslibrary{groupplots,dateplot}
\usetikzlibrary{patterns,shapes.arrows}
\pgfplotsset{compat=newest}
\usepackage{physics}
\usepackage{subcaption}
\usepackage{adjustbox}
\usepackage{amsthm}
\usepackage{tabularx}
\usepackage{multirow}
\usepackage{mathbbol}  
\DeclareSymbolFontAlphabet{\mathbb}{bbold}

\usepackage{fancyhdr}
\newtheorem*{definition}{Definition}

\newcommand{\note}[1]{\textbf{Note:} #1}
\definecolor{Blue1}{HTML}{858585}
\definecolor{Blue2}{HTML}{6F6F6F}
\definecolor{Blue3}{HTML}{555555}
\colorlet{LightBlue}{Blue2!20!white}

\begin{document}

\title{Quantum Optimal Control Using MAGICARP: Combining Pontryagin's Maximum Principle and Gradient Ascent}

\author{Denis Jankovi\'c}
\address{Institut de Physique et Chimie des Mat\'eriaux de Strasbourg UMR-7504 CNRS, Universit\'e de Strasbourg, 23 rue du Loess, Strasbourg, 67000, France}
\address{{Institute of Nanotechnology}, {Karlsruhe Institute of Technology}, Kaiserstra{\ss}e 12, Karlsruhe, 76131, Germany}
\email{denis@jankovic.phd}
\author{Jean-Gabriel Hartmann}
\address{Institut de Physique et Chimie des Mat\'eriaux de Strasbourg UMR-7504 CNRS, Universit\'e de Strasbourg, 23 rue du Loess, Strasbourg, 67000 France}
\address{IPCMS}
\author{Paul-Louis Etienney}
\address{Institut de Physique et Chimie des Mat\'eriaux de Strasbourg UMR-7504 CNRS, Universit\'e de Strasbourg, 23 rue du Loess, Strasbourg, 67000 France}
\address{IPCMS}
\author{Killian Lutz}
\address{Institut de Recherche Mathématique Avancée, UMR 7501 CNRS, Universit\'e de Strasbourg, Inria, 7 Rue René Descartes, Strasbourg, 67000 France}
\address{IRMA}
\author{Yannick Privat}
\address{Institut \'Elie Cartan de Lorraine, CNRS, Inria, Universit\'e de Lorraine, F-54000 Nancy}
\address{IECL}
\address{Institut Universitaire de France}
\author{Paul-Antoine Hervieux}
\address{Institut de Physique et Chimie des Mat\'eriaux de Strasbourg UMR-7504 CNRS, Universit\'e de Strasbourg, 23 rue du Loess, Strasbourg, 67000 France}
\address{IPCMS}
\vspace{10pt}

\date{\today}

\begin{abstract}
We introduce the MAGICARP algorithm, a numerical optimization method for quantum optimal control problems that combines the structure provided by Pontryagin’s Maximum Principle (PMP) and the robustness of gradient ascent techniques, such as GRAPE. MAGICARP is formulated as a "shooting technique," aiming to determine the appropriate initial adjoint momentum to realize a target quantum gate. This method naturally incorporates time and energy optimal constraints through a PMP-informed pulse structure. We demonstrate MAGICARP’s effectiveness through illustrative numerical examples, comparing its performance to GRAPE and highlighting its advantages in specific scenarios.
\end{abstract}

\maketitle

\section{Introduction}
Quantum optimal control (QOC) has emerged as a pivotal technique for manipulating quantum systems efficiently and precisely, essential in quantum computing, quantum sensing, and quantum communication. The primary objective is to design control pulses that drive a quantum system from an initial state to a target state or implement a specific quantum operation with high fidelity, often under stringent constraints on resources such as energy and time.

Two prominent methods in QOC are Gradient Ascent Pulse Engineering (GRAPE)\cite{khaneja_optimal_2005} and Pontryagin’s Maximum Principle (PMP)\cite{pontryagin_mathematical_2017}. GRAPE optimizes discretized control pulses by iteratively adjusting them in the direction of increasing fidelity. Conversely, PMP offers an analytical characterization of optimal solutions but often lacks practical computational implementation, especially for complex quantum systems.

In this work, we introduce MAGICARP (Method for Adjoint- and Gradient-based self-Iterative Construction And Refinement of Pulses), a hybrid method leveraging both the analytical insights from PMP and the numerical efficiency of gradient ascent algorithms. MAGICARP functions as a shooting method in control theory, determining the correct initial adjoint momentum (or costate) required to achieve a desired final state or gate operation. This initial momentum guides the generation of control pulses consistent with the PMP's necessary conditions for optimality, explicitly accommodating either time-optimal or energy-optimal criteria. The time required to implement a gate is a crucial factor, it impacts the coherence time of the system, and the energetic cost of the control Hamiltonians is also an important factor to consider, as it impacts the energy dissipated in the system. Moreover, in the presence of control and environmental errors, the robustness of the gate to these errors is also a key factor to consider. 

In the context of this work we consider two general types of objective or cost functionals \cite{shapiro_lagrange_1966,dalessandro_introduction_2021}, known as problems of 

(i) \textbf{Mayer type}:

\begin{equation}\label{eq:MayerType}
    \mathcal{J}[u] = \phi(\rho(T), T),
\end{equation}

and (ii) \textbf{Lagrange type}:

\begin{equation}\label{eq:LagrangeType}
    \mathcal{J}[u] = \int_0^T \mathcal{L}[\rho(t), u(t), t] \, \mathrm{d}t.
\end{equation}

As shown in \cref{eq:MayerType}, Mayer problems focus on quantities dependent solely on the final state at time $T$, such as maximizing the fidelity between the evolved state $\rho(T)$ and a desired target state. Lagrange problems, represented in \cref{eq:LagrangeType}, account for costs accumulated over the entire time interval $[0, T]$, including factors like the total energy used by control fields $u(t)$ or penalties for deviations during the system's evolution. 

One often considers hybrid cost functionals that combine both Mayer and Lagrange terms, known as problems of \textbf{Bolza type}:

\begin{equation}\label{eq:BolzaType}
    \mathcal{J}[u] = \alpha \phi(\rho(T), T) + (1 - \alpha) \int_0^T \mathcal{L}[\rho(t), u(t), t] \, \mathrm{d}t,
\end{equation}

where $\alpha \in [0, 1]$ is a weighting parameter that balances the optimization priorities. By varying $\alpha$, one can smoothly transition between focusing on the final-state fidelity and minimizing resource costs. For instance, starting with $\alpha = 1$ emphasizes reaching the target state with maximum fidelity. Gradually decreasing $\alpha$ introduces considerations like energy efficiency, allowing the optimization to find control fields that are both effective and resource-conscious \cite{lutz_decoherence_2023}.

The choice of a good cost functional is crucial for the success of the optimization process, in particular if implemented numerically and it is not a global optimization, the wrong choice of cost functional can lead to suboptimal solutions due to the optimization algorithm getting stuck in local minima.

\subsection{Pontryagin's Maximum Principle}

There are numerous numerical optimization algorithms that can be used to solve the optimal control problem, such as the GRAPE algorithm, the Krotov method, or the CRAB algorithm, among others\cite{koch_quantum_2022, glaser_training_2015}. However, these algorithms are often computationally expensive and can be sensitive to the choice of initial conditions and parameters. Moreover, they may not always provide the global optimum of the cost functional, and they may not be able to handle constraints on the control fields.

The cost problem can also be tackled analytically, and, while it is probably not solvable analytically, it can give us insights that could then be used to design better numerical optimization algorithms, with less free parameters for example.

The optimal control problem is a well-studied field in control theory, and one of the most powerful tools to solve it is \textbf{Pontryagin's Maximum Principle (PMP)}. The PMP is a necessary condition for optimality in optimal control problems, and it provides a set of differential equations, known as the Pontryagin equations, that the optimal control fields must satisfy. The PMP is a generalization of the Euler-Lagrange equations in classical mechanics, and it is a cornerstone of optimal control theory.

Let us consider a general optimal control problem, where we aim to minimize a cost functional of the form \cref{eq:LagrangeType}:

\begin{equation}
    \mathcal{J}[x, \vec{u}] = \int_0^T \mathcal{L}[x(t), \vec{u}(t), t] \, \mathrm{d}t,
\end{equation}

where $x(t)$ is the state of the system at time $t$, $\vec{u}(t)$ are the control fields, and $\mathcal{L}[x(t), \vec{u}(t), t]$ is then the Lagrangian of the system. 

Assuming the system is governed by a set of differential equations, known as the state equations, that describe the evolution of the system in the absence of control fields:

\begin{equation}
    \dot{x}(t) = f[x(t), t],
\end{equation}

\begin{definition}
The Pontryagin's Maximum Principle (PMP) states that  if $\vec{u}^*(t)$ is the optimal control field that minimizes the cost functional, then there exists a set of adjoint variables $\vec{\lambda}(t)$ and a pseudo-Hamiltonian $\mathcal{H}[x(t), \vec{\lambda}(t), \vec{u}(t), t]$ such that the optimal control field $\vec{u}^*(t)$ and the optimal state $x^*(t)$ satisfy the following set of equations \cite{Boscain2021}:

\begin{align}
    \mathcal{H}[x, \vec{\lambda}, \vec{u}, t] &= \vec{\lambda}^\dag \cdot f[x, t] + \lambda^0 \mathcal{L}(x,\vec{u},t)\\
    \dot{x} &= \frac{\partial \mathcal{H}}{\partial \vec{\lambda}}, \\
    \dot{\vec{\lambda}}^\dag &= -\frac{\partial \mathcal{H}}{\partial x} = - \vec{\lambda}^\dag \frac{\partial f}{\partial x} -\lambda^0 \frac{\partial \mathcal{L}}{\partial x}, \\
    \vec{0} &= \left.\frac{\partial \mathcal{H}}{\partial \vec{u}}\right\vert_{\vec{u}=\vec{u}^*}, \\
    \mathcal{H}[x^*(t), \vec{\lambda}^*(t), &\vec{u}^*(t), t] = C \geq 0,\quad\forall t\in [0,T].
\end{align}

The optimal control field $\vec{u}^*(t)$ is then given by the solution of the third equation, and the optimal state $x^*(t)$ is given by the solution of the first equation evaluated at the optimal control field.

Here $\lambda^0\leq 0$ is a constant called the abnormal Lagrange multiplier, and $C$ is the constant value of the pseudo-Hamiltonian for the optimal control fields $\vec{u}^*(t)$ over the time interval over which the control problem is defined.

\end{definition}

\note{If $\lambda^0$ is not zero, then it is possible to rescale the equations by defining $\lambda \leftarrow \lambda/|\lambda^0|$, and $\mathcal{H} \rightarrow \mathcal{H}/|\lambda^0|$, such that, effectively $\lambda^0=-1$.}

\note{When one applies the PMP where $x$ are the center of mass coordinates of an object in classical mechanics and the Langrangian is the traditional kinetic energy minus the potential energy, the PMP gives the equations of motion of the object, with $\lambda$ being the momentum $p$ of the object, and the pseudo-Hamiltonian is then the real Hamiltonian, i.e. total energy of the object. Therefore, the PMP is a generalization of the Hamilton equations in classical mechanics.}

In our particular case, the state is not the density matrix (i.e. the litteral state of the system), but the unitary operator $U(t)$, and the control fields are $u_k(t)$ in the control Hamiltonian

\begin{equation}
    H_c(t) = \sum_k u_k(t) H_k(t).
\end{equation}

The previous set of equations can be written as:

\begin{align}
    f[U] &= \dot{U} = -i H U, \label{eq::PMP1}\\
    \mathcal{H}[U, \lambda, \vec{u}, t] &= \lambda^\dag(t) \cdot f[U] - \mathcal{L}[U,\vec{u},t], \label{eq::PMP2}\\
    \dot{\lambda} &= -i H \lambda + \left(\frac{\partial \mathcal{L}}{\partial U}\right)^\dag, \label{eq::PMP3}\\
    \vec{0} &= \left.\frac{\partial \mathcal{H}}{\partial \vec{u}}\right\vert_{u=u^*}, \label{eq::PMP4}\\
    \mathcal{H}[U^*(t), \lambda^*(t), \vec{u}^*(t), t] &= C \geq 0,\quad\forall t\in [0,T].\label{eq::PMP5}
\end{align}

where $\lambda$ is the adjoint matrix, and $\mathcal{L}$ is the Lagrangian to be determined. The scalar product for two matrices $A$ and $B$ is defined as $A\cdot B = \operatorname{ReTr}(A^\dag B)$.

\subsubsection{Multiple control fields for a qudit in the interaction picture}

We apply the Pontryagin Maximum Principle (PMP) to the problem of identifying optimal control fields for implementing a quantum gate on a $d$‑level qudit. Working in the interaction picture, the free evolution is removed and the Hamiltonian is
\begin{equation}
    H(t)=\sum_k u_k(t)H_k,
\end{equation}
where $H_k$ are control Hamiltonians and $u_k(t)$ their corresponding control fields. Our goal is to realise a target unitary $U_{\text{target}}$ at a normalised final time $T=1$.

It will be convienient to define the cumulative envelope of the control fields as
\begin{equation}
    c(t)\equiv\sqrt{\sum_k u_k^2(t)}.
\end{equation}

The choice of the final time $T$ is arbitrary, as, for a driven quantum system in the interaction picture, for any continuous integrable scalar function $S(t)$, the following time-dependent rescaling of the control fields

$$
\Bigl\{u_k(t)\Bigr\}_{k=1}^m 
\;\longrightarrow\;
\Bigl\{\tfrac{u_k(t)}{S(t)}\Bigr\}_{k=1}^m,
$$

accompanied by the time-reparametrisation

$$
t'=\!\!\int_{0}^{t}\!|S(s)|\,ds,
\qquad dt'=|S(t)|\,dt,
\qquad \tilde T=\int_0^T |S(s)|\,ds,
$$

leaves the overall evolution operator $\,U=\mathcal T\,\exp\!\bigl[-\tfrac{i}{\hbar}\!\int_0^{T}\!H(t)\,dt\bigr]=\mathcal T\,\exp\!\bigl[-\tfrac{i}{\hbar}\!\int_0^{\tilde T}\!H(t')\,dt'\bigr]$ unchanged, i.e. the total pulse area is conserved; only the clock that measures it is stretched. 

This means that for a given set of pulses applied over a time interval $[0,1]$, there are easily computable rescaled pulses $u_k(t')=\Omega_\text{max}u_k(t)/c(t)$ that yield the same evolution operator $U(1)$, but with a different duration

\begin{equation}
    \widetilde T=\int_0^1 c(t)/\Omega_\text{max}\,\mathrm dt=\frac{1}{\Omega_\text{max}}\int_0^1 \sqrt{\sum_k u_k^2(t)}\,\mathrm dt,
    \label{eq::T_max}
\end{equation}

Moreover, this rescaling assures that the cumulative envelope $c(t)$ is normalised to a given maximum attainable amplitude constraint on the control fields, $\Omega_\text{max}$, the duration of the gate in units of $\Omega_{\max}^{-1}$ is then simply the integral of the cumulative envelope over the time interval $[0,1]$.

\paragraph{Simple quadratic cost functional}
We first treat the elementary quadratic cost
\begin{equation}
    \mathcal J[\vec u]=\int_0^1\sum_k u_k^2(t)\,\mathrm dt = \int_0^1 c^2(t)\,\mathrm dt,
    \label{eq::costsimple}
\end{equation}
which penalises the integrated pulse energy. PMP gives
\begin{align}
    \mathcal{H}[U(t), \vec{\lambda}(t), \vec{u}(t)] &= \operatorname{ImTr}\left(\lambda^\dag(t) H(t) U(t)\right) - \sum_k u_k^2(t), \label{eq::PMP2Tbis}\\
    \dot{\lambda}(t) &= -i H(t) \lambda(t), \label{eq::PMP3Tbis}\\
    \vec{0} &= \operatorname{ImTr}\left(\lambda^\dag(t) \frac{\partial H(t)}{\partial \vec{u}^*} U(t)\right) - 2 \vec{u}^*(t),\label{eq::PMP4Tbis}\\
    \mathcal{H}[U^*(t), \vec{\lambda}^*(t), \vec{u}^*(t)] &= C \equiv c^2 \geq 0,\quad\forall t\in [0,1].\label{eq::PMP5Tbis}
\end{align}


Let us now analyze the adjoint equation \cref{eq::PMP3Tbis}. Since $U(t) \in SU(d)$, the adjoint state $\lambda(t)$ belongs to the tangent space at $U(t)$ and can be written as:

\begin{equation}
    \lambda(t) = U(t) \lambda_0,
\end{equation}

where $\lambda_0 \in \mathfrak{su}(d)$ is a constant, skew-Hermitian matrix. Inserting into the adjoint equation, we obtain:

\begin{equation}
    \dot{\lambda}_0 = 0 \quad \Rightarrow \quad \lambda(t) = U(t) \lambda_0.
    \label{eq::adjoint_matrix_0}
\end{equation}

Using this in \cref{eq::PMP4Tbis} and the identity $\frac{\partial H}{\partial u_k} = H_k$, we find that for all $k$:

\begin{equation}
    \operatorname{ImTr}\left( U(t) \lambda_0 U^\dag(t) H_k \right) = 2 u_k^*(t).
    \label{eq::optimal_control_l0_2}
\end{equation}

Introducing $g = i \lambda_0$, we find that $g$ is a traceless Hermitian matrix, and the optimal control fields satisfy:

\begin{equation}
    u_k^*(t) = \frac{1}{2} \operatorname{ReTr}\left(U(t) g U^\dag(t) H_k \right) = \frac{1}{2} g(t) \cdot H_k = \frac{1}{2} g \cdot H_k(-t), 
    \label{eq::optimal_control2}
\end{equation}

with $g(t) = U(t) g U^\dag(t)$ the time-evolved adjoint, or equivalently $H_k(-t) = U^\dag(t) H_k U(t)$ the time-reversed Hamiltonians.

$u_k^*(t)$ is uniquely defined by $g(t)$, the latter being a solution of the first-order differential equation:

\begin{equation}
    \dot{g}(t) = -i[H(t), g(t)],
\end{equation}

implies $g(t)$ is uniquely determined by its initial condition $g(0) = g = i \lambda_0$.

Any set of control fields satisfying \cref{eq::optimal_control2} at all times is optimal in the sense of minimizing the cost in \cref{eq::costsimple}, and the pulses -- and \textit{a fortiori} the final gate $U(1)$ -- can be constructed \textit{a priori}, by choosing an initial adjoint momentum $\lambda_0$ or $g$ and then solving the differential equation for $g(t)$, which in turn gives the optimal control fields $u_k^*(t)$. 

\note{The trivial solution $u_k(t) = 0$ corresponds to the identity gate and arises when $g \propto \mathbb{1}_d$, noting that $H_k$ are traceless.}

Moreover, \cref{eq::PMP5Tbis} leads to useful insights about the envelope of the optimal control fields, since, by recalling that $H(t) = \sum_k u_k(t) H_k$, it gives us:

\begin{equation}
    \sum_k u_k^*(t) \operatorname{ImTr}\left(U(t) \lambda_0 U^\dag(t)H_k\right) - \sum_k (u_k^*(t))^2 = c^2,
\end{equation}

which, using \cref{eq::optimal_control_l0_2}, simplifies to:

\begin{equation}
    \sum_k u_k^2(t) = c^2,
\end{equation}

where $c$ is a constant. This means that the cumulative envelope of the control fields is constant over the time interval $[0,1]$, and this implies that for any given maximal driving amplitude $\Omega_{\max}$, as per \cref{eq::T_max}, the duration of the pulse is fixed to 
\begin{equation}
    \widetilde T = \frac{1}{\Omega_{\max}}\int_0^1 \sqrt{\sum_k u_k^2(t)} \, dt = |c|/\Omega_{\max}.
    \label{eq::T_max2}
\end{equation}

Meanwhile, the energy of the pulses is simply 

\begin{equation}
    \widetilde E = \int_0^{\widetilde T} \Omega_{\max}^2 \, dt = |c|\Omega_{\max}.
    \label{eq::E_max}
\end{equation}

Conversely, if the time duration $\widetilde T$ is fixed, the energetic cost of the pulses is 

\begin{equation}
    \widetilde E = c^2 / \widetilde T
    \label{eq::E_max2}
\end{equation}

In both cases, going from the time interval $[0,1]$ to the physical time interval $[0,\widetilde T]$ is done by rescaling the time by a factor $c/\Omega_{\max} = \widetilde T$ and the control fields by a factor $\Omega_{\max}/c=\widetilde T^{-1}$. The latter can be absorbed into $g$ in \cref{eq::optimal_control2} ($\overline{g}_E = g\Omega_{\max}/c = g/T$), which allows to apply the definition of the optimal controls to any time interval $[0,T]$ without loss of generality.

\begin{definition}
For a family of control fields $\{u_k(t)\}$ applied for a time $T$ that is \emph{energy-optimal}—i.e. it minimises the energy cost of the gate, there exists a constant traceless Hermitian matrix $\overline{g}_E$ such that for all $k$ and all $t\in[0,T]$,

\begin{equation}
    u_k(t)=\frac{1}{2}\operatorname{ReTr}\bigl(U(t)\overline{g}_EU^\dagger(t)H_k\bigr),
    \label{eq::optimal_control_2_def}
\end{equation}
for all $k$ and all $t\in[0,T]$, where $U(t)$ is the unitary operator that evolves the system under the control Hamiltonian $H(t)=\sum_k u_k(t)H_k$. The transversality condition of the PMP then imposes that the cumulative envelope $c(t)=\sqrt{\sum_k u_k^2(t)}$ is constant over the time interval $[0,T]$.
\end{definition}

This means that if one wants to implement a gate with a fixed duration $T$, but a minimal energetical cost, a procedure is to find a constant traceless Hermitian matrix $\overline{g}$ such that $U(t)$ is the desired gate at time $t=T$ -- up to a global phase and up to a tolerance on the fidelity of the gate. The energetic cost of the gate, which is simply the integral of the squared cumulative envelope over the time interval $[0,T]$, can be computed from $\overline{g}_E$ as
\begin{equation}
    \widetilde E = \int_0^T \sum_k u_k^2(t)\, dt = \sum_k (u_k^*(0))^2 T = \sum_k \left(\frac{1}{2} \operatorname{ReTr}\left(\overline{g}_E H_k\right)\right)^2 T.
\end{equation}

In the case where the amplitude $\Omega_{\max}$ is fixed, finding a constant traceless Hermitian matrix $g$ such that $U(t)$ is the desired gate at time $t=1$ allows to compute the time necessary to implement the gate while trying to minimize the energy, c.f. \cref{eq::T_max2}, i.e.
\begin{equation}
    \widetilde T = \frac{1}{2\Omega_{\max}} \sqrt{\sum_k  \left( \operatorname{ReTr}\left(g H_k\right) \right)^2}.
\end{equation}

\note{The PMP does not provide a constructive algorithm to find the optimal control fields, as the PMP only gives necessary conditions for optimality, not sufficient ones. However, it does provide a structure that can be used to either construct good candidate pulses, or that cen be incorporated into a numerical optimization algorithm, to improve its convergence toward the optimal solution.}
\paragraph{Time‑optimal cost functional}
To minimise the gate duration itself, we refer to \cref{eq::T_max}, and we take
\begin{equation}
    \mathcal J[\vec u]=\widetilde T=\int_0^1\sqrt{\sum_k u_k^2(t)}\,\mathrm dt = \int_0^1 c(t)\,\mathrm dt,
    \label{eq::Lagrangian}
\end{equation}
PMP now yields

\begin{align}
    \mathcal{H}[U(t), \vec{\lambda}(t), \vec{u}(t)] &= \operatorname{ImTr}\left(\lambda^\dag(t) H(t) U(t)\right) - \sqrt{\sum_k u_k^2(t)}, \label{eq::PMP2T}\\
    \dot{\lambda}(t) &= -i H(t) \lambda(t), \label{eq::PMP3T}\\
    \vec{0} &= \operatorname{ImTr}\left(\lambda^\dag(t) \frac{\partial H(t)}{\partial \vec{u}^*} U(t)\right) - \vec{u}^*(t)/c(t),\label{eq::PMP4T}\\
    \mathcal{H}[U^*(t), \vec{\lambda}^*(t), \vec{u}^*(t)] &= C \geq 0,\quad\forall t\in [0,1].\label{eq::PMP5T}
\end{align}
With $\lambda(t)=U(t)\lambda_0=-iU(t)g$, similarily to the previous cost function under interest, the optimal control fields satisfy
\begin{equation}
    u_k^*(t)=\operatorname{ReTr}\bigl(U(t)gU^\dagger(t)H_k\bigr),
    \label{eq::optimal_control}
\end{equation}
where $g$ is a constant traceless Hermitian matrix, and $c(t)=\sqrt{\sum_k u_k^2(t)}$ is the cumulative envelope of the control fields.

Substituting $H(t)=\sum_k u_kH_k$ and\,(\ref{eq::optimal_control}) into $\mathcal H=C$ (\cref{eq::PMP5T}) gives
\begin{equation}
    \sum_k u_k^*(t)\operatorname{ReTr}\bigl(U(t)gU^\dagger(t)H_k\bigr)-\sqrt{\sum_k u_k^{*2}(t)}=C.
\end{equation}
This leads to the following condition on the cumulative envelope of the control fields:
\begin{equation}
    c(t)(c(t)-1) = C,
\end{equation}
which, assuming c(t) is cotinuous over the time interval $[0,1]$, implies that the cumulative envelope $c(t)$ is also constant over the time interval $[0,1]$.

\note{An intersting consequence, is that for a given maximal amplitude $\Omega_{\max}$, and with an envelope $c(t)$ fixed to the constant $\Omega_{\max}$ over the time interval $[0,\widetilde T]$, the structure of the time-optimal control fields is the same as the energy-optimal pulse in \cref{eq::optimal_control_2_def}, with $\Omega_{\max} \overline{g}_T = \frac{1}{2}\overline{g}_E$.}

\begin{definition}
For a fixed maximum amplitude $\Omega_{\max}$, if a family of control fields $\{u_k(t)\}$ applied for a time $\widetilde T$ is \emph{time-optimal}—that is, it minimises the time required to implement a gate, there exists a constant traceless Hermitian matrix $\overline{g}_T$ such that for all $k$ and all $t\in[0,\widetilde T]$,
\begin{equation}
    \label{eq::optimal_control_time}
    u_k(t)=\Omega_{\max}\,\operatorname{ReTr}\bigl(U(t)\overline{g}_TU^\dagger(t)H_k\bigr).
\end{equation}
\end{definition}

Combining the two necessary conditions for both the time-optimal and energy-optimal control fields, we find that an time-optimal family of control fields $\{u_k(t)\}$ is also energy-optimal over the time interval $[0,\widetilde T]$.

\paragraph{Open questions}
Although the PMP reveals the \emph{structure} of optimal fields, it does not yet provide
\begin{enumerate}[label=\arabic*]
  \item a constructive computational algorithm, nor
  \item a mechanism to incorporate gate fidelity.
\end{enumerate}
These topics are addressed in the next section.

\subsubsection{Single control field for a qudit in the laboratory frame}

Let us now consider the case of a single control field $u(t)$ acting on a qudit system, in the laboratory frame however, where the free evolution is not removed. The system Hamiltonian reads:

\begin{equation}
    H(t) = H_0 + u(t) H_\text{c},
\end{equation}

The free evolution is given by the drift Hamiltonian $H_0$, and the control Hamiltonian is $H_\text{c}$. The goal is to implement a target unitary operator $U_\text{target}$ at time $T$.

The presence of the drift Hamiltonian $H_0$ does not change the form of the PMP equations \cref{eq::PMP2Tbis,eq::PMP3Tbis,eq::PMP4Tbis,eq::PMP5Tbis}, nor does it change the form of the adjoint equation \cref{eq::adjoint_matrix_0}, however the evolution operator $U(t)$ incorporates now the drift Hamiltonian.

\Cref{eq::optimal_control_2_def} as a necessary condition for energy-optimality is therefore still valid with the drift Hamiltonian included in the evolution operator $U(t)$, and the optimal control fields are given by:
\begin{equation}
    u^*(t) = {\Omega_\text{max}} \operatorname{ReTr}\left(U(t) \overline{g}_E U^\dag(t) H_\text{c} \right),
\end{equation}

where $\overline{g}_E $ is a constant, traceless, Hermitian matrix. 

The first major difference with the interaction picture is that rescaling the duration of the gate to $T=1$ does not imply that the duration of the maximally driven control field is equal to its energetic cost as in \cref{eq::T_max} in units of $\Omega_\text{max}^{-1}$. Therefore, a pulse of the aforementioned would only be energy-optimal for a fixed duration \textit{e.g.} $T=1$, and not time-optimal. However, the case of energy-optimal control fields is still interesting, and the last PMP equation 

\begin{equation}
    \mathcal{H}[U^*(t), \vec{\lambda}^*(t), u^*(t)] = C \geq 0,\quad\forall t\in [0,1],
\end{equation}

allows us to determine that the envelope of the control field in the presence of a drift Hamiltonian is given by:

\begin{equation}
    c(t) = |u^*(t)| = \sqrt{C-\operatorname{ReTr}(U(t) \overline{g}_E U^\dag(t) H_0)}.
\end{equation}

This leads to the following constrained differential equation for $\overline{g}_E(t) = U(t) \overline{g}_E U^\dag(t)$ to satisfy:

\begin{equation}
\begin{cases}
    \dot{\overline{g}_E }(t) = -i[H_0 + \operatorname{ReTr}(\overline{g}_E (t) H_\text{c})H_\text{c}, \overline{g}(t)], \\
    \left(C - \operatorname{ReTr}(\overline{g}_E (t) H_0)\right)^2 = \operatorname{ReTr}(\overline{g}_E (t) H_\text{c})^2.
\end{cases}
\end{equation}

The case of time-optimal control fields with a drift Hamiltonian is more complex, as no rescaling of the time interval is possible, the time cost cannot be expressed in terms of the cumulative envelope, the correct cost functional is actually given by the control-independent time integral:

\begin{equation}
    \mathcal J[T]=\int_0^T \mathrm dt=T,
    \label{eq::cost_time_optimal}
\end{equation}

this means that the Langrangian $\mathcal L$ is simply $1$, moreover since $T$ is not fixed anymore, the PMP equations are given by \cite{Boscain2021}:

\begin{align}
    \mathcal{H}[U(t), \lambda(t), u(t)] &= \operatorname{ImTr}\left(\lambda^\dag(t) H(t) U(t)\right) - 1, \label{eq::PMP2T_time_optimal}\\
    \dot{\lambda}(t) &= -i H(t) \lambda(t), \label{eq::PMP3T_time_optimal}\\
    0 &= \operatorname{ImTr}\left(\lambda^\dag(t) \frac{\partial H(t)}{\partial u^*} U(t)\right),\label{eq::PMP4T_time_optimal}\\
    \mathcal{H}[U^*(t), \lambda^*(t), u^*(t)] &= 0,\quad\forall t\in [0,T].\label{eq::PMP5T_time_optimal}
\end{align}

The adjoint equation is unchanged, and we can rewrite the transversality condition \cref{eq::PMP5T_time_optimal} as:
\begin{equation}
    \operatorname{ReTr}\left(U(t) g U^\dag(t) H_0\right) + u^*(t) \operatorname{ReTr}\left(U(t) g U^\dag(t) H_\text{c}\right) = 1,
\end{equation}

where $g = i \lambda_0$ is a constant, traceless, Hermitian matrix. We find that the time-optimal control fields satisfy:

\begin{equation}
    u^*(t) = \frac{1 - \operatorname{ReTr}\left(U(t) g U^\dag(t) H_0\right)}{\operatorname{ReTr}\left(U(t) g U^\dag(t) H_\text{c}\right)}.
\end{equation}

Or, equivalently, by defining $\Phi(t) = \operatorname{ReTr}\left(U(t) g U^\dag(t) H_\text{c}\right)$, we have
\begin{equation}
    u^*(t) = \frac{1 - \operatorname{ReTr}\left(U(t) g U^\dag(t) H_0\right)}{\Phi(t)} = \frac{1 - \operatorname{ReTr}\left(U(t) g U^\dag(t) H_0\right)}{\|\Phi(t)\|} \frac{\Phi(t)}{\|\Phi(t)\|},
\end{equation}

If constraints are imposed on the maximal amplitude of the control field, the optimal control field is \textit{bang-bang}, i.e. it switches between -$\Omega_{\max}$ and $\Omega_{\max}$ \cite{Boscain2021}, except on singular arcs where $\operatorname{ReTr}\left(U(t) g U^\dag(t) H_\text{c}\right)=0$, which do generally not persist, and 

\begin{equation}
    \left|\frac{1 - \operatorname{ReTr}\left(U(t) g U^\dag(t) H_0\right)}{\operatorname{ReTr}\left(U(t) g U^\dag(t) H_\text{c}\right)}\right| = \Omega_{\max}.
\end{equation}

Moreover if $[H_0, H_\text{c}] = 0$, $\dot \Phi(t) = 0$, and the control field is constant, which is a trivial case. 

However, in the unbounded case, the only physically acceptable solution is on singular arcs, otherwise the control field diverges.

This imposes the following constraints on $g(t) = U(t) g U^\dag(t)$:
\begin{equation}
\begin{cases}
    \dot{g}(t) = -i[H_0 + u^*(t)H_\text{c}, g(t)], \\
    \operatorname{ReTr}(g(t) H_\text{c}) = 0, \\
    \operatorname{ReTr}(g(t) H_0) = 1.
\end{cases}
\end{equation}

On a singular arc we have $\Phi(t)\equiv 0$. Differentiating once gives
\begin{align}
\dot\Phi(t)
&= \operatorname{ReTr}\!\big(\dot g(t)\,H_{\mathrm c}\big)
= \operatorname{ReTr}\!\big(-i[H_0+uH_{\mathrm c},\,g(t)]\,H_{\mathrm c}\big) \nonumber\\
&= \operatorname{ReTr}\!\big(g(t)\,i[H_0+uH_{\mathrm c},\,H_{\mathrm c}]\big)
= \operatorname{ReTr}\!\big(g(t)\,i[H_0,\,H_{\mathrm c}]\big),
\end{align}
so the singularity persists only if $\dot\Phi(t)=0$, i.e.
\[
\operatorname{ReTr}\!\big(g(t)\,i[H_0,\,H_{\mathrm c}]\big)=0.
\]
Differentiating again (note that $H_0$ and $H_{\mathrm c}$ are time-independent),
\begin{align}
\ddot\Phi(t)
&= \operatorname{ReTr}\!\big(\dot g(t)\,i[H_0,\,H_{\mathrm c}]\big)
= \operatorname{ReTr}\!\big(-i[H_0+uH_{\mathrm c},\,g(t)]\,i[H_0,\,H_{\mathrm c}]\big) \nonumber\\
&= \operatorname{ReTr}\!\big(g(t)\,i[H_0+uH_{\mathrm c},\,i[H_0,\,H_{\mathrm c}]]\big) \nonumber\\
&= \operatorname{ReTr}\!\big(g(t)\,i[H_0,\,i[H_0,\,H_{\mathrm c}]]\big)
\;+\;u(t)\,\operatorname{ReTr}\!\big(g(t)\,i[H_{\mathrm c},\,i[H_0,\,H_{\mathrm c}]]\big).
\end{align}
Enforcing $\ddot\Phi(t)=0$ on the singular arc yields the (second-order) singular control:
\begin{equation}
u^*(t)
= -\,\frac{\operatorname{ReTr}\!\big(g(t)\,[H_0,\,[H_0,\,H_{\mathrm c}]]\big)}
{\operatorname{ReTr}\!\big(g(t)\,[H_{\mathrm c},\,[H_0,\,H_{\mathrm c}]]\big)}
\label{eq:singular_u_second_order}
\end{equation}
provided the denominator is nonzero. If $[H_{\mathrm c},\,[H_0,\,H_{\mathrm c}]] = 0$, one must differentiate $\Phi(t)$ further until the control appears (higher-order singular arc).

\section{MAGICARP algorithm}
MAGICARP iteratively refines an adjoint matrix $g$ to optimize the target quantum operation in the interaction picture. The algorithm proceeds as follows:
\begin{enumerate}
\item Initialize adjoint matrix $g$.
\item Compute initial controls using:
\begin{equation}
u_k(0) = \frac{1}{2}\operatorname{ReTr}(g H_k).
\end{equation}
\item Compute the unitary evolution using discretized pulses:
\begin{equation}
U(\delta t) = \exp\left(-i \delta t \sum_k u_k(0) H_k\right).
\end{equation}
\item Update control fields at subsequent steps using adjoint propagation:
\begin{equation}
\tilde{u}_k(\delta t) = \frac{1}{2} \operatorname{ReTr}\left(U(\delta t) g U^\dag(\delta t) H_k \right).
\end{equation}
\item Iterate until final time $T$.
\item Evaluate fidelity and update $g$ using gradient ascent until convergence.
\end{enumerate}

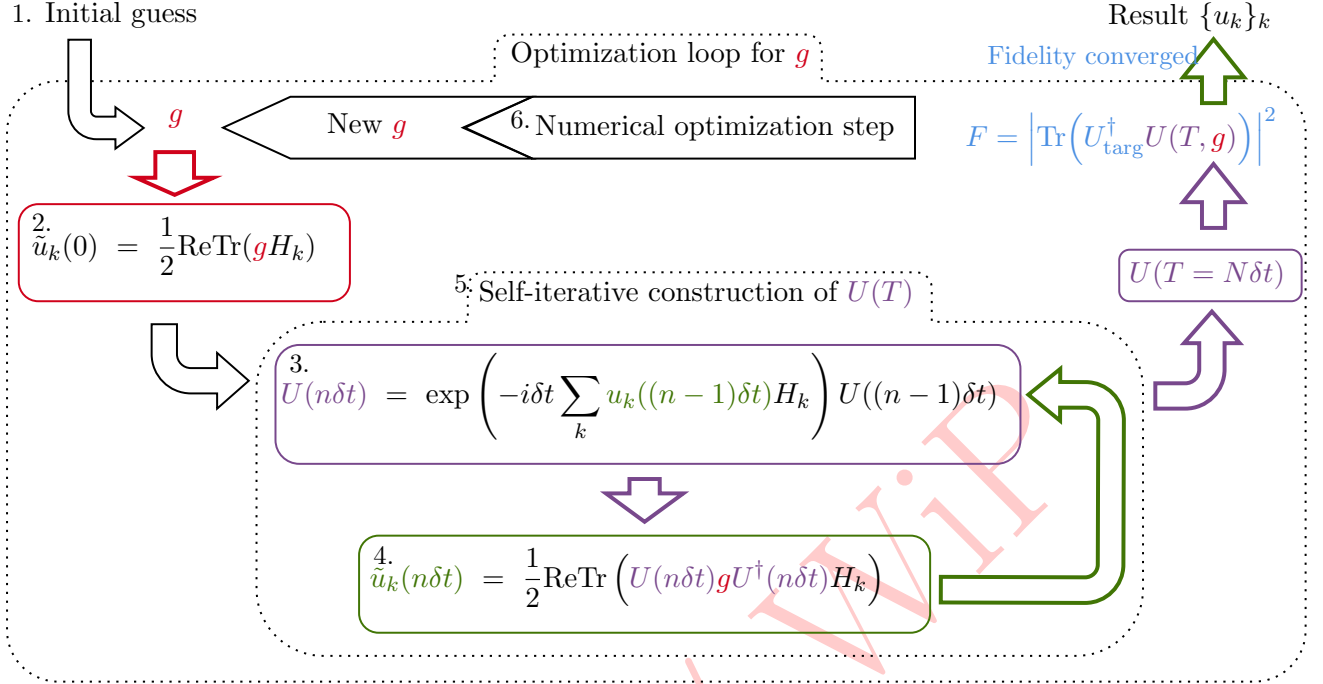
\begin{figure}[htb!]
\centering
\tikzset{every picture/.style={line width=0.75pt}} 

\begin{tikzpicture}[x=0.75pt,y=0.75pt,yscale=-1,xscale=1]

\draw  [color={rgb, 255:red, 208; green, 2; blue, 27 }  ,draw opacity=1 ] (9,120.4) .. controls (9,114.99) and (13.39,110.6) .. (18.8,110.6) -- (166.2,110.6) .. controls (171.61,110.6) and (176,114.99) .. (176,120.4) -- (176,149.8) .. controls (176,155.21) and (171.61,159.6) .. (166.2,159.6) -- (18.8,159.6) .. controls (13.39,159.6) and (9,155.21) .. (9,149.8) -- cycle ;

\draw   (76,171.2) -- (76,186.32) .. controls (76,197.26) and (84.87,206.13) .. (95.82,206.13) -- (109.03,206.13) -- (109.03,213.8) -- (125,199.74) -- (109.03,185.68) -- (109.03,193.35) -- (95.82,193.35) .. controls (91.93,193.35) and (88.78,190.2) .. (88.78,186.32) -- (88.78,171.2) -- cycle ;
\draw  [color={rgb, 255:red, 120; green, 72; blue, 140 }  ,draw opacity=1 ][line width=1.5]  (582.2,213.8) -- (597.32,213.8) .. controls (608.26,213.8) and (617.13,204.93) .. (617.13,193.98) -- (617.13,180.77) -- (624.8,180.77) -- (610.74,164.8) -- (596.68,180.77) -- (604.35,180.77) -- (604.35,193.98) .. controls (604.35,197.87) and (601.2,201.02) .. (597.32,201.02) -- (582.2,201.02) -- cycle ;
\draw  [color={rgb, 255:red, 120; green, 72; blue, 140 }  ,draw opacity=1 ] (564,138.8) .. controls (564,135.93) and (566.33,133.6) .. (569.2,133.6) -- (650.8,133.6) .. controls (653.67,133.6) and (656,135.93) .. (656,138.8) -- (656,154.4) .. controls (656,157.27) and (653.67,159.6) .. (650.8,159.6) -- (569.2,159.6) .. controls (566.33,159.6) and (564,157.27) .. (564,154.4) -- cycle ;

\draw  [color={rgb, 255:red, 120; green, 72; blue, 140 }  ,draw opacity=1 ] (138.47,193.67) .. controls (138.47,187.05) and (143.85,181.67) .. (150.47,181.67) -- (502.18,181.67) .. controls (508.8,181.67) and (514.18,187.05) .. (514.18,193.67) -- (514.18,229.67) .. controls (514.18,236.3) and (508.8,241.67) .. (502.18,241.67) -- (150.47,241.67) .. controls (143.85,241.67) and (138.47,236.3) .. (138.47,229.67) -- cycle ;
\draw  [color={rgb, 255:red, 120; green, 72; blue, 140 }  ,draw opacity=1 ][line width=1.5]  (305.13,256.6) -- (317,256.6) -- (317,249.33) -- (332.25,249.33) -- (332.25,256.6) -- (344.13,256.6) -- (324.63,270.92) -- cycle ;
\draw  [color={rgb, 255:red, 65; green, 117; blue, 5 }  ,draw opacity=1 ] (180.63,287.72) .. controls (180.63,282.31) and (185.01,277.92) .. (190.43,277.92) -- (458.83,277.92) .. controls (464.24,277.92) and (468.63,282.31) .. (468.63,287.72) -- (468.63,317.12) .. controls (468.63,322.54) and (464.24,326.92) .. (458.83,326.92) -- (190.43,326.92) .. controls (185.01,326.92) and (180.63,322.54) .. (180.63,317.12) -- cycle ;

\draw  [color={rgb, 255:red, 65; green, 117; blue, 5 }  ,draw opacity=1 ][line width=1.5]  (474,311.64) -- (546.73,311.88) .. controls (557.39,311.92) and (566.06,303.3) .. (566.1,292.64) -- (566.34,219.78) .. controls (566.38,209.12) and (557.76,200.45) .. (547.1,200.42) -- (534.23,200.37) -- (534.26,193.76) -- (518.18,206.78) -- (534.17,219.91) -- (534.19,213.3) -- (547.06,213.34) .. controls (550.58,213.35) and (553.43,216.22) .. (553.42,219.74) -- (553.18,292.6) .. controls (553.16,296.12) and (550.3,298.97) .. (546.78,298.96) -- (474.04,298.71) -- cycle ;
\draw  [color={rgb, 255:red, 0; green, 0; blue, 0 }  ,draw opacity=1 ][dash pattern={on 0.84pt off 2.51pt}][line width=0.75]  (162.62,170) -- (235.45,170) -- (235.45,152.92) .. controls (235.45,148.55) and (239,145) .. (243.37,145) -- (461.53,145) .. controls (465.9,145) and (469.45,148.55) .. (469.45,152.92) -- (469.45,170) -- (542.28,170) .. controls (560.9,170) and (576,185.1) .. (576,203.72) -- (576,304.88) .. controls (576,323.5) and (560.9,338.6) .. (542.28,338.6) -- (162.62,338.6) .. controls (144,338.6) and (128.9,323.5) .. (128.9,304.88) -- (128.9,203.72) .. controls (128.9,185.1) and (144,170) .. (162.62,170) -- cycle ;
\draw  [color={rgb, 255:red, 120; green, 72; blue, 140 }  ,draw opacity=1 ][line width=1.5]  (627,109.6) -- (617.4,109.6) -- (617.4,122.6) -- (605.73,122.6) -- (605.73,109.6) -- (596.13,109.6) -- (611.56,90.33) -- cycle ;
\draw   (461,87.6) -- (268,87.6) -- (233,72) -- (268,56.4) -- (461,56.4) -- cycle ;
\draw  [color={rgb, 255:red, 208; green, 2; blue, 27 }  ,draw opacity=1 ][line width=1.5]  (68.13,97.28) -- (77.88,97.28) -- (77.88,84.33) -- (97.38,84.33) -- (97.38,97.28) -- (107.13,97.28) -- (87.63,105.92) -- cycle ;
\draw   (268,56.4) -- (146,56.4) -- (112,72) -- (146,87.6) -- (268,87.6) -- (234,72) -- cycle ;
\draw  [color={rgb, 255:red, 0; green, 0; blue, 0 }  ,draw opacity=1 ][dash pattern={on 0.84pt off 2.51pt}][line width=0.75]  (249,32.92) .. controls (249,28.55) and (252.55,25) .. (256.92,25) -- (406.08,25) .. controls (410.45,25) and (414,28.55) .. (414,32.92) -- (414,48.52) -- (632.08,48.52) .. controls (646.4,48.52) and (658,60.12) .. (658,74.44) -- (658,79.52) -- (658,79.52) -- (658,325.68) .. controls (658,340) and (646.4,351.6) .. (632.08,351.6) -- (29,351.6) .. controls (14.68,351.6) and (3.08,340) .. (3.08,325.68) -- (3.08,74.44) .. controls (3.08,60.12) and (14.68,48.52) .. (29,48.52) -- (249,48.52) -- (249,32.92) -- cycle ;
\draw  [fill={rgb, 255:red, 255; green, 255; blue, 255 }  ,fill opacity=1 ] (34,27.6) -- (34,63.58) .. controls (34,71.45) and (40.38,77.83) .. (48.25,77.83) -- (57.75,77.83) -- (57.75,82.6) -- (72,72.13) -- (57.75,61.66) -- (57.75,66.43) -- (48.25,66.43) .. controls (46.68,66.43) and (45.4,65.15) .. (45.4,63.58) -- (45.4,27.6) -- cycle ;
\draw  [color={rgb, 255:red, 65; green, 117; blue, 5 }  ,draw opacity=1 ][fill={rgb, 255:red, 255; green, 255; blue, 255 }  ,fill opacity=1 ][line width=1.5]  (627,46.6) -- (617.4,46.6) -- (617.4,59.6) -- (605.73,59.6) -- (605.73,46.6) -- (596.13,46.6) -- (611.56,27.33) -- cycle ;

\draw (14,115.6) node [anchor=north west][inner sep=0.75pt]   [align=left] {$\displaystyle \tilde{u}_{k}( 0) \ =\ \frac{1}{2}\mathrm{ReTr}(\textcolor[rgb]{0.82,0.01,0.11}{g} H_{k})$};
\draw (184.63,282.92) node [anchor=north west][inner sep=0.75pt]   [align=left] {$\displaystyle \textcolor[rgb]{0.25,0.46,0.02}{\tilde{u}_{k}( n\delta t)} \ =\ \frac{1}{2}\mathrm{ReTr}\left(\textcolor[rgb]{0.47,0.28,0.55}{U( n\delta t)}\textcolor[rgb]{0.82,0.01,0.11}{g}\textcolor[rgb]{0.47,0.28,0.55}{U^{\dagger }( n\delta t)} H_{k}\right)$};
\draw (567.5,137.6) node [anchor=north west][inner sep=0.75pt]   [align=left] {$\displaystyle \textcolor[rgb]{0.47,0.28,0.55}{U( T=N\delta t)}$};
\draw (139.63,183.92) node [anchor=north west][inner sep=0.75pt]   [align=left] {$\displaystyle \textcolor[rgb]{0.47,0.28,0.55}{U( n\delta t)} \ =\ \exp\left( -i\delta t\sum _{k}\textcolor[rgb]{0.25,0.46,0.02}{u_{k}(( n-1) \delta t)} H_{k}\right) U(( n-1) \delta t)$};
\draw (485,60.4) node [anchor=north west][inner sep=0.75pt]  [color={rgb, 255:red, 208; green, 2; blue, 27 }  ,opacity=1 ]  {\textcolor[rgb]{0.29,0.56,0.89}{$F=\left| \Tr(U_{\mathrm{targ}}^{\dagger }\textcolor[rgb]{0.47,0.28,0.55}{U(T, \textcolor[rgb]{0.82,0.01,0.11}{g})})\right| ^{2}$}};
\draw (83,61.4) node [anchor=north west][inner sep=0.75pt]    {$\textcolor[rgb]{0.82,0.01,0.11}{g}$};
\draw (268,64) node [anchor=north west][inner sep=0.75pt]   [align=left] {Numerical optimization step};
\draw (163,63) node [anchor=north west][inner sep=0.75pt]   [align=left] {New $\displaystyle \textcolor[rgb]{0.82,0.01,0.11}{g}$};
\draw (239.45,147) node [anchor=north west][inner sep=0.75pt]   [align=left] {Self-iterative construction of $\displaystyle \textcolor[rgb]{0.47,0.28,0.55}{U( T)}$};
\draw (4,7) node [anchor=north west][inner sep=0.75pt]   [align=left] {{\footnotesize 1.} Initial guess};
\draw (557,7) node [anchor=north west][inner sep=0.75pt]   [align=left] {Result $\displaystyle \{u_{k}\}_{k}$};
\draw (496,29) node [anchor=north west][inner sep=0.75pt]   [align=left] {{\footnotesize \textcolor[rgb]{0.29,0.56,0.89}{Fidelity converged}}};
\draw (255.5,27) node [anchor=north west][inner sep=0.75pt]   [align=left] {Optimization loop for $\displaystyle \textcolor[rgb]{0.82,0.01,0.11}{g}$};
\draw (14,114) node [anchor=north west][inner sep=0.75pt]   [align=left] {{\footnotesize 2.}};
\draw (144,185) node [anchor=north west][inner sep=0.75pt]   [align=left] {{\footnotesize 3.}};
\draw (186.43,281.92) node [anchor=north west][inner sep=0.75pt]   [align=left] {{\footnotesize 4.}};
\draw (227,144.92) node [anchor=north west][inner sep=0.75pt]   [align=left] {{\footnotesize 5.}};
\draw (255.43,61.92) node [anchor=north west][inner sep=0.75pt]   [align=left] {{\footnotesize 6.}};

\end{tikzpicture}
\caption{Flowchart of the MAGICARP algorithm.
\label{fig::flowMAGICARP}}
\end{figure}

\Cref{fig::flowMAGICARP} shows the flowchart of the MAGICARP algorithm.

Let us compare the numerical requirements and advantages of MAGICARP and GRAPE, the two algorithms that we have discussed in this chapter. 


\paragraph{Number of optimization parameters}

GRAPE requires $N_{\text{steps}}\times N_{\text{controls}}$ real optimization parameters, where $N_{\text{steps}}$ is the number of time steps for the discretization of the evolution and $N_{\text{controls}}$ is the number of control fields. MAGICARP on the other hand requires $d^2-1$ real optimization parameters, where $d$ is the dimension of the system. This is because MAGICARP optimizes the adjoint matrix $g$, which is a $d\times d$ traceless hermitian matrix, and the control fields are then computed from $g$ through the self-iterative method. The dimension of the optimization space can then be very different for the two methods, and is summarised in \cref{tab::optimization_parameters}.

\begin{table}[htb!]
    \centering
    \begin{tabular}{|c|c|c|}
        \hline
        \rowcolor{LightBlue} \textbf{Algorithm} &\textbf{ \# of optimization parameters} \\
        \hline
        \cellcolor{LightBlue} \textbf{GRAPE} & $N_{\text{steps}}\times N_{\text{controls}}$ \\
        \cellcolor{LightBlue} \textbf{MAGICARP} & $d^2-1$ \\
        \hline
    \end{tabular}
    \caption{Comparison of the number of optimization parameters required by GRAPE and MAGICARP.}
    \label{tab::optimization_parameters}
\end{table}

MAGICARP can be very useful for systems with a large number of control fields, and/or a large number of time steps, it also yields, by construction, more continuous-looking pulses. However, MAGICARP scales poorly with the dimension of the system, as the number of optimization parameters grows quadratically with the dimension of the system, on the other hand GRAPE scales linearly with the number of control fields and time steps, which, for systems with a small number of control fields relatively to the dimension of the system, can yield a smaller number of optimization parameters than MAGICARP, in particular as the number of time steps can always be adjusted and therefore allows for a trade-off between the number of optimization parameters and the accuracy of the optimization for example.


\paragraph{Numerical complexity and stability}

The numerical complexity of the two algorithms is also different. GRAPE relies on analytically computed gradients of the cost functional with respect to the control fields, which can be evaluated efficiently. MAGICARP on the other hand relies on the numerical computation of the gradients of the cost functional with respect to the adjoint matrix $g$, which can be computationally expensive. 

An interesting point can be made about the numerical stability of the MAGICARP method, as the adjoint matrix $g$ is a traceless hermitian matrix, it is possible to use the Lie algebra $\mathfrak{su}(d)$ to represent $g$ as a linear combination of the generators (up to a global prefactor $i$) of $SU(d)$, which could maybe be used to compute the gradients of the cost functional with respect to $g$ efficiently.

Moreover, the dependence of the final evolution operator $U(N\delta)$ with $N$ the number of time steps, on the initial guess for the adjoint matrix $g$ can be a source of numerical instability. $U(N\delta) \in SU(d)$, therefore it has $d^2-1$ real degrees of freedom, and the adjoint matrix $g$ has also $d^2-1$ real degrees of freedom, which means that the optimization space is the same for the two matrices. However, the mapping between the two spaces is not straightforward and highly non-linear, and an in-depth study of the stability of $U(N\delta t)$ or at least $\Tr(U^\dag_\text{targ}U(N\delta t))$ with respect to small variations in $g$ is necessary. For example, can two different adjoint matrices $g_1$ and $g_2$ that are very close in the optimization space lead to two very different final evolution operators $U(N\delta)$? 

Finally a crucial point is the stability and convergence of the solution with respect to $\delta t$, the time step for the discretization of the evolution. Does the solution converge to the optimal continuous solution for $\delta t\rightarrow 0$? A possible improvement to the algorithm could be to use a variable time step, where the time step is decreased as the optimization progresses, in order to refine the solution for example. Otherwise, the discretization of the evolution can be accounted for in the PMP in the first place for example, as in the work of Dionis \textit{et al.} \cite{dionis_time-optimal_2023}.

\paragraph{Optimality of the solution}
A straightforward advantage of MAGICARP is that it combines the insights from the PMP to restrict the mathematical structure of the optimal control fields with respect to a Lagrange type cost functional. The gradient-ascent is then used to adjust this mathematical form in order to minimize the Mayer-type cost functional that is the fidelity of the obtained gate. Therefore, MAGICARP aims for an time or energy-optimal solution, while basic GRAPE only aims for a solution that minimizes the fidelity, moreover if one includes other costs in GRAPE, the gradient of the new cost functional with respect to the control fields amplitude at each time step is not straightforward to compute, and the optimization process can be very slow. 

\subsubsection{Example of results}

\begin{figure}[htb!]
    \centering
    \hfill
    \begin{subfigure}[t]{0.48\textwidth}
        \includegraphics[width=\textwidth, trim={1cm 1cm 1cm 1cm}, clip]{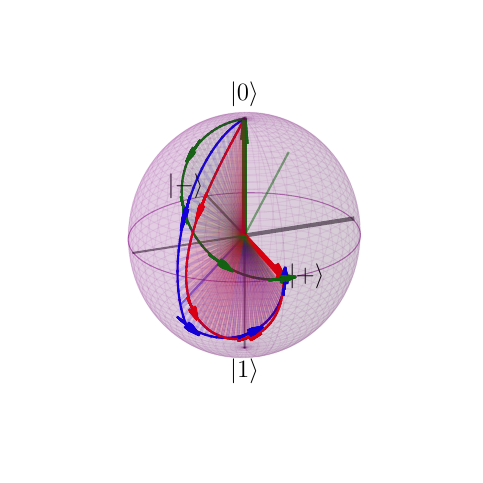}
        \caption{Bloch sphere evolution of the two-qubit system under (green) the optimal control field, (blue) the constrained control field and (red) the MAGICARP constrained control field. The initial state is $\ket{0}$ and represented by a green vector, the target state is $\ket{+}$ and represented by a red vector.}
    \end{subfigure}%
    \hfill%
    \begin{subfigure}[t]{0.48\textwidth}
        \begin{adjustbox}{max width=\columnwidth}
        \input{ch4_fig/control_pulses_magicarp_norm.tex}
        \end{adjustbox}
        \caption{Control amplitudes $f_x(t)$ and $f_y(t)$ in units of $\tau_\text{QSL}^{-1}$ for (gray) the non-optimal constrained Hamiltonian and (color) the MAGICARP constrained control field.}
    \end{subfigure}
    \caption{Comparision of the MAGICARP constrained control field with the constrained one from \cref{fig::bloch_sphere_control_pulses} for the Hadamard gate. a) represents the evolution on the Bloch sphere of the state $\ket{0}$ under (green) the optimal control field $H = \frac{\pi}{2\sqrt{\pi}}\left(\sigma_x + \sigma_y\right)$, (blue) the constrained control field from \cref{fig::bloch_sphere_control_pulses} and (red) the MAGICARP constrained control field $H(t) = f_x(t)\sigma_x + f_y(t)\sigma_y$. b) represents the control amplitudes $f_x(t)$ and $f_y(t)$ in units of $\tau_\text{QSL}^{-1}$ for the constrained control field (gray) and the MAGICARP constrained control field (color). One notices that the MAGICARP constrained control field is more continuous and has a duration of $\sim1.25\tau_{QSL}$ compared to the constrained control field from \cref{fig::bloch_sphere_control_pulses} that has a duration of $\sim1.33\tau_{QSL}$.
    \label{fig::bloch_sphere_control_pulses_MAGICARP}}
\end{figure}

As an example, consider the two-qubit system and the Hadamard gate $H = \frac{1}{\sqrt{2}}\begin{pmatrix} 1 & 1 \\ 1 & -1 \end{pmatrix}$ as the target gate, the control Hamiltonians are chosen to be $\sigma_x$ and $\sigma_y$. The results of the optimization process for the Hadamard gate using MAGICARP are shown in \cref{fig::bloch_sphere_control_pulses_MAGICARP} where it is compared with the constrained control field from \cref{fig::bloch_sphere_control_pulses}. The MAGICARP constrained control field is more continuous and has a duration of $\sim1.25\tau_{QSL}$ compared to the constrained control field from \cref{fig::bloch_sphere_control_pulses} that has a duration of $\sim1.33\tau_{QSL}$. The latter was actually calculated using an augmented GRAPE algorithm with 2 time-steps, with the term $$\left(1-\int_0^T \left(\sum_k u_k^2\right)dt\tau_\text{QSL}^{-1}\right)^2$$ added to the cost functional $1-\Tr(U_{\text{targ}}^\dag U(T))$ to ensure that the duration of the gate is minimized.

As an example, let us consider the optimization of the Hadamard gate for systems of dimension $d = 2, 3, 4, 5, 6$. For each dimension, the control Hamiltonians in the interaction picture are chosen to be the $2(d-1)$ generalized Pauli matrices $\sigma^x_{k,k+1}$ and $\sigma^y_{k,k+1}$
\footnote{We recall that $\sigma^x_{k,k+1} = \ket{k}\bra{k+1} + \ket{k+1}\bra{k}$ and $\sigma^y_{k,k+1} = -i\ket{k}\bra{k+1} + i\ket{k+1}\bra{k}$, where $\ket{k}$ is the $k$-th computational basis state.}. The adjoint matrix $g$ is initialized randomly, and the optimization process is repeated 300 times for each dimension.

\begin{figure}[htb!]
    \centering
    \begin{adjustbox}{max width=\textwidth}
        \input{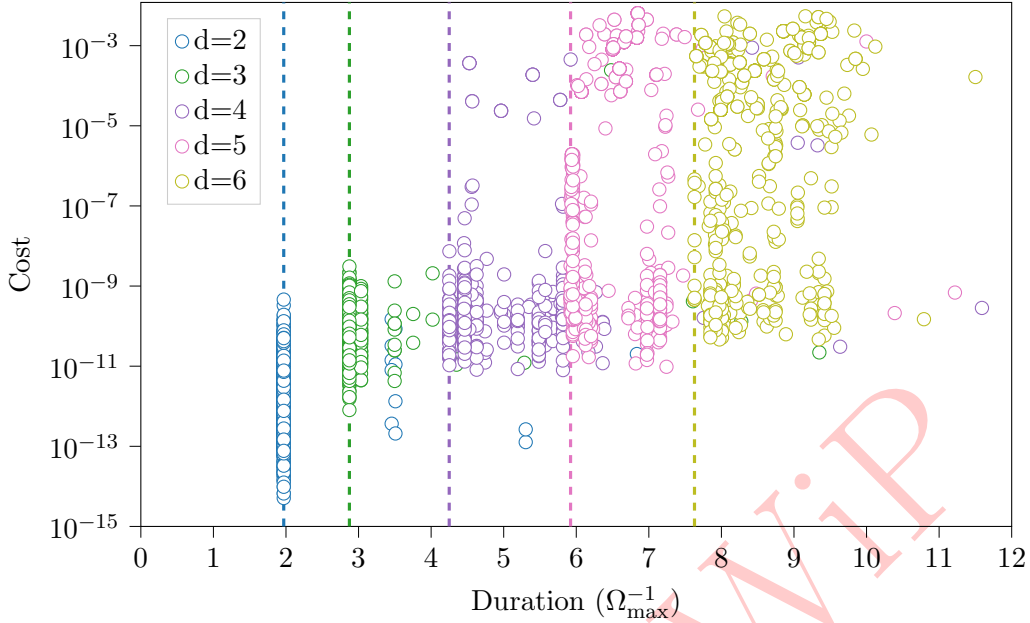}
    \end{adjustbox}
    \caption{Results of 600 runs of the MAGICARP algorithm with the Hadamard gate as a target with random initial guesses for the adjoint matrix $g$. Each color represents a different dimension $d$ of the system $d=2,3,4,5,6$, there is then $3000$ runs in total. The cost functional $1-\frac{1}{d}\Tr(U_{\text{targ}}^\dag U(T))$ is plotted as a function of the duration of the gate ($T$) in units of $\Omega_{\text{max}}^{-1}$, where $\Omega_{\text{max}}$ is the Rabi fields maximally driving the system. 
    \label{fig::cost_duration_qu2-6it-Chrestenson}}
\end{figure}

\begin{figure}[htb!]
    \centering
    \begin{adjustbox}{max width=\textwidth}
        \input{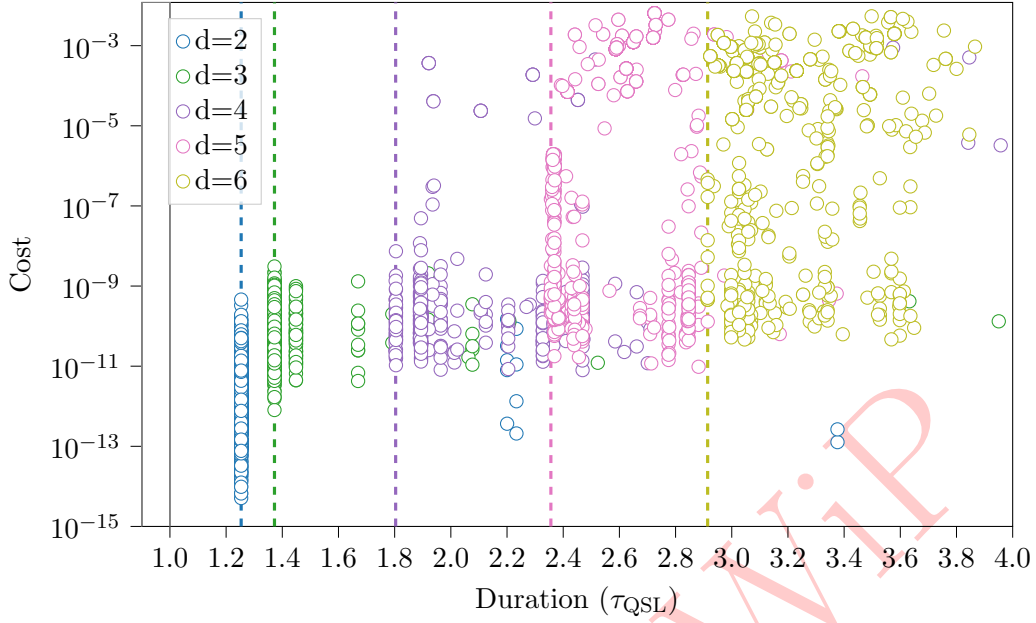}
    \end{adjustbox}
    \caption{Same as \cref{fig::cost_duration_qu2-6it-Chrestenson} but the durations are given in units of $\tau_{QSL}$, the quantum speed limit time. The dashed lines represent, for eached dimension, the minimal duration reached with a cost functional $1-\frac{1}{d}\Tr(U_{\text{targ}}^\dag U(T))$ of $10^{-7}$ or lower.
    \label{fig::cost_duration_qu2-6it-Chrestenson_LG}}
\end{figure}

The results of this optimization process for the QFT gate using MAGICARP are shown in \cref{fig::cost_duration_qu2-6it-Chrestenson}, where the cost functional $1-\frac{1}{d}\Tr(U_{\text{targ}}^\dag U(T))$ is plotted as a function of the gate duration $T$ in units of $\Omega_{\text{max}}^{-1}$, where $\Omega_{\text{max}}$ represents the maximum Rabi field driving the system. The results are also presented in \cref{fig::cost_duration_qu2-6it-Chrestenson_LG}, where the durations are given in units of $\tau_{QSL}=\pi/\Omega_\text{max}\left(1-\frac{1}{d}\right)$, the quantum speed limit time for the QFT($d$) gate.

The results indicate that as the dimension increases, the fidelity of achieving the target gate decreases, and the required duration increases. This is expected because the optimization space grows quadratically with the system's dimension. Consequently, the numerical optimization is more likely to converge to local minima, rather than the global minimum, as the dimension increases. This also explains why, at higher dimensions, the optimization results are more scattered and the process becomes less stable.

It is noteworthy however that, for every dimension, there appears to be a minimal achievable duration for the gate. The dashed lines in \cref{fig::cost_duration_qu2-6it-Chrestenson,fig::cost_duration_qu2-6it-Chrestenson_LG} indicate, for each dimension, the minimal duration at which a cost functional of $1-\frac{1}{d}\Tr(U_{\text{targ}}^\dag U(T))$ of $10^{-7}$ or lower is reached. A significant fraction of the runs achieve durations close to this minimal duration, particularly for lower dimensions. This suggests that the optimization process is often able to find solutions that are near-optimal and possibly close to the quantum speed limit time with the chosen constrained set of control fields. The latter is different than the previously discussed $\tau_\text{QSL}$, the quantum speed limit with unconstrained set of controls.

If verified, it would be interesting to analyze how this minimal duration evolves with the system's dimension, especially in comparison to the quantum speed limit time with unconstrained control fields. As it stands, the minimal duration appears to deviate further from the quantum speed limit time with unconstrained control fields as the system's dimension increases. This can be expected since, while the set of control Hamiltonians satisfies the Lie Algerba Rank Condition (LARC), the ratio of these control Hamiltonians to the total number of $SU(d)$ generators decreases as the system's dimension increases. Specifically, this ratio is $2(d-1)/(d^2-1) = 2/(d+1)$.


\subsubsection{Conclusive remarks}

The MAGICARP algorithm seems to provide some insights on the quantum speed limit with a constrained set of control pulses. In particular, in a system where the control hamiltonians are the $2(d-1)$ generalized Pauli-X and Pauli-Y matrices between two adjacent levels of the system, the minimal achievable durations seem to increase with the dimension, and to stray from optimality relative to an unconstrained set of linearly independent $d^2-1$ control Hamiltonians as the dimension increases. This would imply that in order to minimize the duration of a gate with increased dimension, a higher connectivity is required, i.e. more allowed and controllable transitions.

The results also suggest that the optimization process is often able to find solutions that are near-optimal and possibly close to the quantum speed limit time with the chosen constrained set of control fields however. The MAGICARP algorithm proves quite promising, and, provided the necessary improvements and studies f the numerical stability and convergence are made, it could be a valuable tool for the optimization of quantum gates in the future.

\bibliography{references_natbib}

\end{document}

%% file: ch4_fig/control_pulses_magicarp_norm.tex
\begin{tikzpicture}

    \definecolor{darkgray176}{RGB}{176,176,176}
    \definecolor{lightgray204}{RGB}{204,204,204}
    \definecolor{steelblue31119180}{RGB}{31,119,180}

\begin{groupplot}[group style={group size=1 by 2, x descriptions at=edge bottom, vertical sep=.75em}]
    \nextgroupplot[
      width=\columnwidth,
      height=.6\columnwidth,
      font=\small,
  legend cell align={left},
  legend style={
    fill opacity=0.8,
    draw opacity=1,
    text opacity=1,
    at={(0.03,0.97)},
    anchor=north west,
    draw=lightgray204
  },
  tick align=inside,
  tick pos=left,
  x grid style={darkgray176},
  xtick={-0.2,0,0.2,0.4,0.6,0.8,1,1.2,1.4},
  xmin=-0.0807741183570704, xmax=1.39903201537105,
  y grid style={darkgray176},
  ymin=0.7, ymax=1.717,
  ytick style={color=black}
  ]
  \addplot [ultra thick, gray, const plot mark left, forget plot]
  table {%
  0 1.2605044682104
  0.656843361502751 1.30347987855284
  1.33241144321052 1.30347987855284
  };
  \addplot [ultra thick, Blue1]
  table {%
0 0.793473087287043
0.0125264988311296 0.822421012689867
0.0250558907832617 0.850967168238027
0.0375881720452047 0.879099294670357
0.050123338807691 0.906805373000085
0.0626613872633859 0.934073628907915
0.0752023136068968 0.960892536986884
0.0877461140347816 0.987250824838087
0.100292784745558 1.01313747701651
0.112842321939713 1.03854173882625
0.125394721819709 1.06345311996452
0.137949980589998 1.08786139801388
0.150508094457024 1.11175662178228
0.163069059629236 1.13512911449047
0.175632872317096 1.1579694768066
0.188199528733086 1.18026858972766
0.200769025091718 1.20201761730775
0.213341357609544 1.22320800923305
0.225916522505162 1.24383150324354
0.238494515999225 1.26388012740155
0.251075334314452 1.28334620220738
0.263658973675632 1.30222234256216
0.276245430309637 1.32050145957829
0.288834700445429 1.33817676223793
0.301426780314067 1.35524175889991
0.314021666148717 1.37169025865564
0.326619354184659 1.38751637253473
0.339219840659296 1.40271451456073
0.351823121812164 1.41727940265812
0.364429193884936 1.43120605941098
0.377038053121436 1.44448981267447
0.389649695767642 1.45712629604001
0.402264118071695 1.4691114491551
0.414881316283912 1.48044151789885
0.427501286656786 1.49111305441449
0.440124025445004 1.50112291699986
0.452749528905444 1.51046826985723
0.465377793297191 1.51914658270371
0.478008814881544 1.52715563024368
0.490642589922021 1.5344934915045
0.503279114684367 1.5411585490372
0.515918385436567 1.54714948798335
0.528560398448847 1.55246529501001
0.541205149993687 1.5571052571141
0.553852636345825 1.56106896029804
0.566502853782268 1.56435628811825
0.579155798582298 1.56696742010833
0.59181146702748 1.56890283007864
0.60446985540167 1.57016328429425
0.617130959991022 1.57074983953301
0.629794777083996 1.57066384102567
0.642461302971367 1.56990692028011
0.655130533946228 1.56848099279152
0.667802466304003 1.56638825564072
0.680477096342453 1.56363118498253
0.693154420361681 1.56021253342637
0.70583443466414 1.55613532731119
0.718517135554645 1.55140286387686
0.731202519340373 1.54601870833427
0.743890582330878 1.53998669083623
0.756581320838091 1.53331090335155
0.769274731176333 1.5259956964444
0.781970809662319 1.51804567596133
0.794669552615167 1.50946569962826
0.807370956356404 1.50026087355964
0.820075017209974 1.49043654868229
0.832781731502245 1.47999831707612
0.845491095562015 1.46895200823418
0.85820310572052 1.45730368524441
0.870917758311442 1.44505964089547
0.883635049670914 1.43222639370912
0.896354976137529 1.41881068390145
0.909077534052344 1.4048194692755
0.921802719758892 1.39025992104771
0.934530529603183 1.37513941961049
0.947260959933715 1.35946555023359
0.959994007101479 1.34324609870653
0.972729667459966 1.32648904692461
0.985467937365174 1.30920256842102
0.998208813175616 1.2913950238474
1.01095229125232 1.27307495640538
1.02369836795886 1.2542510872315
1.03644703966131 1.23493231073809
1.04919830272832 1.2151276899123
1.06195215353106 1.19484645157607
1.07470858844327 1.17409798160913
1.08746760384125 1.15289182013769
1.10022919610385 1.13123765669122
1.11299336161251 1.10914532532955
1.12576009675125 1.08662479974293
1.13852939790666 1.06368618832729
1.15130126146794 1.04033972923708
1.16407568382688 1.01659578541813
1.17685266137787 0.992464839622801
1.18963219051792 0.967957489409814
1.20241426764666 0.943084442131006
1.21519888916634 0.917856509907389
1.22798605148184 0.89228460459674
1.24077575100066 0.866379732754991
};
\addlegendentry{$f_x$}
    
    \nextgroupplot[
      width=\columnwidth,
      height=.6\columnwidth,
      font=\small,
    legend cell align={left},
    legend style={fill opacity=0.8, draw opacity=1, text opacity=1, at={(0.03,0.03)},anchor=south west,
    draw=lightgray204},
    tick align=inside,
    tick pos=left,
    x grid style={darkgray176},
    xlabel={$t/\tau_\text{QSL}$},
    xmin=-0.0666205721605262, xmax=1.39903201537105,
    xtick style={color=black},
    y grid style={darkgray176},
    ymin=-1.5, ymax=1.5,
    ytick style={color=black}
    ]
    \addplot [ultra thick, gray, const plot mark left, forget plot]
    table {%
    0 0.937299090949067
    0.656843361502751 -0.876550800855387
    1.33241144321052 -0.876550800855387
    };
    \addplot [ultra thick, Blue1!50!red]
table {%
0 1.35565539869965
0.0125264988311296 1.33829173880679
0.0250558907832617 1.32032419460271
0.0375881720452047 1.30176247079965
0.050123338807691 1.28261651157722
0.0626613872633859 1.26289649459136
0.0752023136068968 1.24261282491179
0.0877461140347816 1.2217761288914
0.100292784745558 1.20039724797125
0.112842321939713 1.17848723242473
0.125394721819709 1.15605733504445
0.137949980589998 1.13311900477559
0.150508094457024 1.10968388029907
0.163069059629236 1.08576378356824
0.175632872317096 1.06137071330266
0.188199528733086 1.0365168384424
0.200769025091718 1.01121449156652
0.213341357609544 0.985476162279158
0.225916522505162 0.959314490566709
0.238494515999225 0.932742260129661
0.251075334314452 0.905772391692433
0.263658973675632 0.87841793629471
0.276245430309637 0.850692068567666
0.288834700445429 0.822608079998452
0.301426780314067 0.794179372186301
0.314021666148717 0.765419450093577
0.326619354184659 0.736341915295073
0.339219840659296 0.706960459228794
0.351823121812164 0.677288856451494
0.364429193884936 0.647340957902124
0.377038053121436 0.617130684176385
0.389649695767642 0.586672018815493
0.402264118071695 0.555979001612248
0.414881316283912 0.525065721937456
0.427501286656786 0.49394631208972
0.440124025445004 0.462634940671549
0.452749528905444 0.431145805994735
0.465377793297191 0.399493129517864
0.478008814881544 0.367691149318803
0.490642589922021 0.335754113604966
0.503279114684367 0.30369627426409
0.515918385436567 0.271531880458245
0.528560398448847 0.239275172263707
0.541205149993687 0.20694037435932
0.553852636345825 0.174541689765893
0.566502853782268 0.142093293639137
0.579155798582298 0.109609327118608
0.59181146702748 0.077103891235045
0.60446985540167 0.044591040878472
0.617130959991022 0.0120847788293524
0.629794777083996 -0.0204009501449508
0.642461302971367 -0.0528522651262183
0.655130533946228 -0.0852553548117789
0.667802466304003 -0.117596483209196
0.680477096342453 -0.149861995257193
0.693154420361681 -0.182038322370861
0.70583443466414 -0.214111987909212
0.718517135554645 -0.246069612563225
0.731202519340373 -0.277897919662574
0.743890582330878 -0.30958374039929
0.756581320838091 -0.341114018966654
0.769274731176333 -0.372475817611703
0.781970809662319 -0.403656321599748
0.794669552615167 -0.434642844089394
0.807370956356404 -0.465422830916595
0.820075017209974 -0.495983865286331
0.832781731502245 -0.526313672370569
0.845491095562015 -0.556400123811195
0.85820310572052 -0.5862312421267
0.870917758311442 -0.615795205021435
0.883635049670914 -0.645080349596315
0.896354976137529 -0.6740751764599
0.909077534052344 -0.702768353738864
0.921802719758892 -0.731148720986885
0.934530529603183 -0.759205292991074
0.947260959933715 -0.7869272634751
0.959994007101479 -0.814304008698239
0.972729667459966 -0.841325090949611
0.985467937365174 -0.867980261936949
0.998208813175616 -0.894259466069275
1.01095229125232 -0.920152843632943
1.02369836795886 -0.945650733860517
1.03644703966131 -0.970743677892073
1.04919830272832 -0.995422421628489
1.06195215353106 -1.01967791847642
1.07470858844327 -1.04350133198464
1.08746760384125 -1.06688403837153
1.10022919610385 -1.08981762894353
1.11299336161251 -1.11229391240443
1.12576009675125 -1.13430491705536
1.13852939790666 -1.15584289288558
1.15130126146794 -1.1769003135539
1.16407568382688 -1.19746987826101
1.17685266137787 -1.21754451351268
1.18963219051792 -1.23711737477403
1.20241426764666 -1.25618184801516
1.21519888916634 -1.27473155114831
1.22798605148184 -1.29276033535686
1.24077575100066 -1.31026228631665
};
\addlegendentry{$f_y$}
\end{groupplot}

\end{tikzpicture}